\documentclass[%
 aip,
 amsmath,amssymb,
 reprint,%
]{revtex4-2}

\usepackage{needspace}
\usepackage[utf8]{inputenc}
\usepackage[T1]{fontenc}
\usepackage{mathptmx}
\usepackage{graphicx}
\usepackage{dcolumn}
\usepackage{bm}
\usepackage{booktabs}
\usepackage{array}
\usepackage{xcolor}
\usepackage{listings}
\usepackage{microtype}
\usepackage{url}
\usepackage[colorlinks=true,
            linkcolor=blue!50!black,
            citecolor=blue!50!black,
            urlcolor=blue!50!black]{hyperref}

\newcommand{\C}{\mathbb{C}}
\newcommand{\Hmat}{\hat H}

\begin{document}

\title{Memory-Scalable and Hardware-Adaptive Matrix-Free Quantum Simulation}

\author{Uriel Shafir}
\affiliation{The Institute of Chemistry, The Hebrew University of Jerusalem, Jerusalem 9190401, Israel}

\author{Ronnie Kosloff}
\affiliation{The Institute of Chemistry, The Hebrew University of Jerusalem, Jerusalem 9190401, Israel}

\date{\today}

\begin{abstract}
The core step in quantum simulations is typically matrix vector multiplication $\phi = \Hmat \psi$. Executing this step is limited by memory requirement to store the Hamiltonian. We present a memory-scalable, hardware-adaptive matrix-free framework for applying large operators on vectors without materializing the full matrix on a single accelerator.

The operator is represented through a block-procedural interface: blocks may
be generated, loaded, cached, distributed, or applied directly only when their
action is needed. For quantum simulation, it provides the core kernel for quantum operations.
An adaptive planner selects block size, cache strategy, GPU grouping, row
distribution, and task parallelization from memory and workload estimates. We describe analytic, measured, 
and learned planning strategies that choose
between procedural generation, partial caching, full caching, and row-distributed caching.
The method removes the requirement that the full dense matrix fit in the accelerator memory. This shifts large simulations from a fixed memory barrier
to a tunable balance between block generation, cache reuse, data movement,
parallel scheduling, and numerical accuracy.
\end{abstract}

\maketitle

\section{Introduction}

Exact and near-exact numerical quantum simulations play an important role in
modern quantum science and engineering. They are used to benchmark quantum devices, validate
quantum algorithms, study many-body dynamics, test approximations, and provide
reference calculations for quantum technologies\cite{de2026universal}. As quantum hardware improves,
simulation on classical hardware maintains its importance; rather, the need for
larger and more reliable reference simulations becomes more acute.

At the computational level, the method developed in this work addresses a more
general primitive: the application of a large linear operator or matrix to a
vector,
\begin{equation}
\vec y = {\bf A} \vec x ,
\end{equation}
where $(A\in \mathbb{C} ^{M\times D}), (x\in \mathbb{C} ^{D})$ and
$(y\in \mathbb{C} ^{M})$. The matrix (${\bf A}$) need not be square. Thus, the same
block-procedural and matrix-free ideas apply to general rectangular
matrix-vector products, linear maps, filters, projections, embeddings,
superoperators, and other large-scale linear transformations. The central
requirement is not that (${\bf A}$) be a Hamiltonian, but that the action (${\bf A} \vec x$) can
be evaluated from blocks, generated entries, cached pieces, or fused local
actions without explicitly storing the full matrix.

Quantum numerical simulation provides a prominent and physically important
special case. Taking $({\bf A}=\Hmat), (\vec x=\psi)$, and $(\vec y=\phi)$ defines the
Hamiltonian-vector product
\begin{equation}
\label{eq:2}
\phi = \Hmat \psi .
\end{equation}
This operation appears is the basic step in any polynomial approximation of a function of $f(\Hmat)$
\cite{kosloff1994propagation}.
For example a Chebychev polynomial expansion of the real-time propagation $e^{-i \Hmat t}$ \cite{tal1984accurate}, imaginary-time filtering \cite{kosloff1986direct,wall1995extraction,baer1997chebyshev},
Krylov methods \cite{eckseler2025escaping}, Green functions \cite{braun2014numerical} and more \cite{saad1986gmres}. 
Typically in these
quantum-simulation algorithms, repeated application of $\Hmat$ is the
dominant computational kernel \cite{westerhout2023implementing}.

Although the discussion in this paper is framed around quantum simulation, the
same computational strategy is relevant beyond quantum mechanics \cite{saad1986gmres,saad1989numerical,de1987product,dax2017new,baglama2005augmented}. Many
large-scale problems in scientific computing and artificial intelligence are
dominated by repeated applications of large linear maps, often too large to
store or move efficiently in explicit dense form. Matrix-free, block-generated,
cached, and hardware-adaptive matrix-vector multiplication can therefore be
useful wherever the main bottleneck is the application of a large matrix or
operator rather than the formal definition of the matrix itself \cite{charlier2021kernel}.

If the Hilbert-space dimension is \(D\), explicit storage of a dense
Hamiltonian requires \(O(D^2)\) complex numbers. This quadratic storage cost is
the matrix-size problem addressed in this work. The vector itself has only
linear size \(O(D)\). Although vector storage can also be important in very
large simulations, it is not the focus of the present framework. Here we focus
on avoiding the need to materialize the full \(D\times D\) Hamiltonian matrix
on one accelerator.

The Hamiltonian is not stored as a global matrix. Instead, it is represented
by a procedural interface that can produce dense block entries, or their
action, when they are needed. The global Hamiltonian remains a well-defined
matrix, but the computation never requires materializing it as a full
\(D\times D\) object on one device.

The framework is general and can be applied to Hamiltonians whose blocks are
generated from reproducible random seeds, reconstructed from analytic
formulas, loaded from host memory or external storage, retrieved from a
database, or evaluated through a fused action. These sources are treated
uniformly by the matrix-free interface. The important requirement is that, for
each pair of block indices \((n,m)\), the framework can obtain the block
\(\Hmat^{[n,m]}\) or compute the product \(\Hmat^{[n,m]}\psi^{[m]}\)
directly.

A fixed block representation is still not enough. The efficient choice of
block size, caching strategy, data layout, and parallelization depends on the
hardware. A calculation that is best served by full matrix caching on a
large-memory GPU may require row-distributed full caching, partial caching, or
procedural generation on a smaller system. A scan over coupling strengths may
parallelize naturally over parameter values, while a single large application
may require splitting the output block rows across multiple GPUs \cite{wietek2018sublattice}. We therefore
include a hardware-adaptive execution layer as part of the method.

The examples in this paper focus on two broad tasks. First, we describe
propagation using polynomial, Krylov, or stepwise integrators whose dominant
kernel is repeated application of \(\Hmat\). Second, we describe stochastic
thermal pure quantum calculations, in which a thermal state is represented by
a filtered random vector and observables are estimated from one or more
typical pure states\cite{sugiura2013canonical}. In both cases, the computational engine is the same
blockwise matrix-free Hamiltonian-vector product.

The paper is organized as follows. Section~\ref{sec:model} defines the block
representation. Section~\ref{sec:matvec} presents the procedural matrix-free
matrix-vector product. Section~\ref{sec:adaptive} describes hardware-adaptive
execution planning, including analytic, measured, and learned plan-selection
algorithms. Section~\ref{sec:memory} analyzes matrix-storage and
runtime scaling. Section~\ref{sec:gpu} discusses GPU-resident caching and row
distribution. Section~\ref{sec:propagation} presents propagation use cases. Section~\ref{sec:limitations} discusses limitations and design implications. Section~\ref{sec:discussion} concludes with scope, extensions, and limitations.

\section{Hamiltonian Block Representation}
\label{sec:model}

Let \(\mathcal{H}\) be a finite-dimensional Hilbert space of dimension \(D\).
A quantum state in this Hilbert space is represented by a vector
\[
    \psi \in \C^D,
\]
and the Hamiltonian is represented by a linear operator
\[
    \Hmat:\C^D\rightarrow \C^D .
\]
In a standard dense-matrix representation, \(\Hmat\) would be stored as a
\(D\times D\) complex matrix. For large \(D\), this representation is usually
impossible on a single GPU because the memory required to store \(\Hmat\)
scales as \(D^2\). The purpose of the block representation introduced here is
not to change the mathematical Hamiltonian, but to group its matrix elements
and vector components in a way that is useful for computation.

We partition the full vector space into \(N_{\mathrm{blk}}\) computational
blocks. The size of block \(n\) is denoted by \(b_n\), so that
\[
    D = \sum_{n=0}^{N_{\mathrm{blk}}-1} b_n .
\]
Here \(N_{\mathrm{blk}}\) is the number of blocks, while \(b_n\) is the number
of vector components stored in the \(n\)-th block. The blocks may all have the
same size, but this is not required. The partition is a computational device:
it does not necessarily correspond to a physical decomposition into
subsystems, lattice sites, energy sectors, symmetry sectors, or system-bath
components. Such interpretations may exist in specific applications, but they
are not assumed by the framework.

The block number \(N_{\mathrm{blk}}\) should therefore be understood as an
execution parameter, not as a physical parameter. Changing \(N_{\mathrm{blk}}\)
changes how the matrix is partitioned, but should not change the mathematical
operator being applied. In seeded or procedural implementations this requires
care: if the random generator is indexed by block labels rather than by global
matrix indices, then changing the block partition can produce a different
random Hamiltonian realization. A partition-invariant implementation should
generate matrix elements, or deterministic block entries, from global row and
column indices. Otherwise, \(N_{\mathrm{blk}}\) becomes part of the ensemble
definition rather than a purely computational choice.

With this partition, the state vector is written as
\[
\psi =
\begin{pmatrix}
\psi^{[0]} \\
\psi^{[1]} \\
\vdots \\
\psi^{[N_{\mathrm{blk}}-1]}
\end{pmatrix},
\qquad
\psi^{[n]} \in \C^{b_n} .
\]
The notation \(\psi^{[n]}\) denotes the \(n\)-th block of the state vector. If
\(\alpha\) labels a component inside block \(n\), then
\[
    \psi^{[n]}_\alpha,
    \qquad
    \alpha = 0,\ldots,b_n-1,
\]
denotes the \(\alpha\)-th component of the \(n\)-th block. The pair
\((n,\alpha)\) therefore labels one component of the original global vector
\(\psi\).

The Hamiltonian is partitioned in the corresponding way. It is written as a
block matrix
\begin{equation}
\Hmat =
\begin{pmatrix}
\Hmat^{[0,0]} & \Hmat^{[0,1]} & \cdots & \Hmat^{[0,N_{\mathrm{blk}}-1]} \\
\Hmat^{[1,0]} & \Hmat^{[1,1]} & \cdots & \Hmat^{[1,N_{\mathrm{blk}}-1]} \\
\vdots & \vdots & \ddots & \vdots \\
\Hmat^{[N_{\mathrm{blk}}-1,0]} &
\Hmat^{[N_{\mathrm{blk}}-1,1]} &
\cdots &
\Hmat^{[N_{\mathrm{blk}}-1,N_{\mathrm{blk}}-1]}
\end{pmatrix} .
\label{eq:blockham}
\end{equation}
The block \(\Hmat^{[n,m]}\) maps input block \(m\) of the vector into output
block \(n\). Therefore,
\[
    \Hmat^{[n,m]} \in \C^{b_n\times b_m} .
\]
The first index \(n\) refers to the output block, and the second index \(m\)
refers to the input block. In component notation,
\[
    \Hmat^{[n,m]}_{\alpha\beta}
\]
is the matrix element that couples component \(\beta\) of input block \(m\) to
component \(\alpha\) of output block \(n\).

It is important to emphasize that Eq.~\eqref{eq:blockham} is completely
general. Every finite-dimensional Hamiltonian can be written in this form after
choosing a partition of the vector components. The block representation does
not imply that the Hamiltonian is block diagonal. It also does not imply that
most blocks are zero. In the most general case, every block
\(\Hmat^{[n,m]}\) may be dense and nonzero. Thus, the framework allows the
Hamiltonian to be fully dense at the global level, even though the notation
groups the matrix entries into blocks.

The action of the Hamiltonian on a state vector is
\[
    \phi = \Hmat\psi ,
\]
where \(\phi\in\C^D\) is the output vector. After partitioning \(\phi\) in the
same way as \(\psi\),
\[
\phi =
\begin{pmatrix}
\phi^{[0]} \\
\phi^{[1]} \\
\vdots \\
\phi^{[N_{\mathrm{blk}}-1]}
\end{pmatrix},
\qquad
\phi^{[n]} \in \C^{b_n},
\]
the matrix-vector product becomes
\begin{equation}
\phi^{[n]}
=
\sum_{m=0}^{N_{\mathrm{blk}}-1}
\Hmat^{[n,m]}\psi^{[m]} .
\label{eq:blockmatvec}
\end{equation}
In component form,
\begin{equation}
\phi^{[n]}_{\alpha}
=
\sum_{m=0}^{N_{\mathrm{blk}}-1}
\sum_{\beta=0}^{b_m-1}
\Hmat^{[n,m]}_{\alpha\beta}
\psi^{[m]}_{\beta},
\qquad
\alpha = 0,\ldots,b_n-1 .
\label{eq:blockmatvecindex}
\end{equation}
This expression is the ordinary dense matrix-vector product written with two
levels of indexing: a block index and an index inside each block.

For clarity, consider the special case in which all blocks have the same size,
\[
    b_n=b,
    \qquad
    D=N_{\mathrm{blk}}b .
\]
Then every block \(\Hmat^{[n,m]}\) is a \(b\times b\) matrix, and the full
Hamiltonian contains \(N_{\mathrm{blk}}^2\) such blocks. If all blocks are
dense and nonzero, storing all of them is equivalent to storing a dense
\(D\times D\) Hamiltonian. Therefore, the block notation by itself does not
reduce the number of mathematical matrix elements. It only specifies how the
entries of the Hamiltonian and the components of the vector are grouped.

No approximation has been introduced by this notation. The Hamiltonian remains
the full operator \(\Hmat\), and the block decomposition only specifies how its
matrix elements are indexed. The computational advantage comes in the next
step: instead of assembling all blocks at once, the matrix-vector product can
be evaluated by generating, loading, caching, or applying only the blocks
required for the current part of the computation.

\section{Procedural Matrix-Free Matrix-Vector Multiplication}
\label{sec:matvec}

The block representation of Sec.~\ref{sec:model} is useful because it allows
the product
\[
    \phi = \Hmat\psi
\]
to be evaluated without assembling the full \(D\times D\) Hamiltonian in GPU
memory. Section~\ref{sec:model} defined the mathematical object. This section
describes the computational procedure used to apply that object.

For a single pair of block indices \((n,m)\), we represent this operation
abstractly as
\begin{equation}
    \texttt{apply\_H\_block}(n,m,\psi^{[m]})
    =
    \mathcal{G}(n,m;\Theta_{n,m})\,\psi^{[m]} .
    \label{eq:single_block_apply}
\end{equation}
Here \(\mathcal{G}(n,m;\Theta_{n,m})\) denotes the procedural representation
of the Hamiltonian block \(\Hmat^{[n,m]}\). The symbol \(\Theta_{n,m}\) denotes
the data required to define that block. Depending on the implementation,
\(\Theta_{n,m}\) may contain a random seed, parameters of an analytic formula,
a database key, a file offset, compression metadata, cached block data, or
other information needed to reconstruct or apply the block.

Equation~\eqref{eq:single_block_apply} is an interface, not necessarily an
instruction to explicitly build the block as a dense array. In the most direct
implementation, the block \(\Hmat^{[n,m]}\) is generated or loaded as a dense
array, multiplied by \(\psi^{[m]}\), and then released. More memory-restricted
implementations may compute the same block action in smaller internal pieces,
or may fuse block generation with the contraction. The external operation
remains the block product \(\Hmat^{[n,m]}\psi^{[m]}\).

With this interface, the complete output block is
\begin{equation}
    \phi^{[n]}
    =
    \sum_{m=0}^{N_{\mathrm{blk}}-1}
    \texttt{apply\_H\_block}(n,m,\psi^{[m]}) .
    \label{eq:output_block_apply}
\end{equation}
Thus, the full matrix-vector product can be interpreted as a loop over output
blocks \(n\), where each output block is formed by a loop over input blocks
\(m\). The global Hamiltonian is never materialized. Instead, the algorithm
evaluates block contributions and accumulates them into the output vector.

\subsection{Sources and storage modes for block data}

The same matrix-free interface supports several modes of specifying and
storing Hamiltonian blocks.

\paragraph{Seeded generation.}
A block may be generated reproducibly from a seed:
\begin{equation}
    \Hmat^{[n,m]} = \mathcal{G}_{\mathrm{seed}}(n,m,s_{n,m}) .
\end{equation}
The stored object is the seed \(s_{n,m}\) and the generation rule, not the
dense matrix itself. This is useful for random-matrix benchmarks, stochastic
environments, or controlled ensembles of Hamiltonians.

\paragraph{Analytic generation.}
A block may be defined by a formula
\begin{equation}
    \Hmat^{[n,m]}_{\alpha\beta}
    =
    f(n,m,\alpha,\beta;\theta).
\end{equation}
The matrix entries are computed when needed. In this case the Hamiltonian is
defined by the function \(f\) and the parameter set \(\theta\), rather than by
a stored dense array.

\paragraph{Externally stored blocks.}
A block may be stored outside GPU memory and retrieved by key:
\begin{equation}
    \Hmat^{[n,m]}
    =
    \mathcal{D}[\mathrm{key}(n,m)] .
\end{equation}
The external storage can contain complete dense blocks, compressed
representations, deterministic model records, calibrated experimental blocks,
or learned matrix surrogates. The computational kernel remains unchanged: the
block is loaded or reconstructed only for the currently active part of the
matrix-vector product.

\paragraph{Fused action.}
The most memory-efficient implementation may expose only the action
\begin{equation}
    \psi^{[m]} \mapsto \Hmat^{[n,m]}\psi^{[m]}
\end{equation}
without ever forming \(\Hmat^{[n,m]}\) as a dense array. This is still the same
Hamiltonian-vector product and can be used by any algorithm that requires only
applications of \(\Hmat\).

\paragraph{Full caching.}
If memory permits, all required blocks can be stored on the GPU. This removes
repeated generation or loading costs and is advantageous when the same
Hamiltonian is applied many times, as in Krylov, Chebyshev, and TPQ
calculations.

\paragraph{Partial caching.}
If full caching does not fit in memory, a fixed memory budget can be used to
cache only a subset of blocks. Cached blocks are reused directly, while
uncached blocks are generated or loaded on demand. For Hermitian operators, it
is natural to cache canonical block pairs \((n,m)\) with \(n\leq m\), using
\[
    \Hmat^{[m,n]} = \left(\Hmat^{[n,m]}\right)^\dagger .
\]
Partial caching interpolates between full procedural regeneration and full
matrix caching.

\subsection{Block-row accumulation}

For dense Hamiltonians, every output block generally receives contributions
from every input block. The exact deterministic product therefore includes all
block pairs. The calculation may still be scheduled by output block rows: a
GPU or GPU group can own a subset of output blocks and accumulate
\[
    \phi^{[n]}
    =
    \sum_{m=0}^{N_{\mathrm{blk}}-1}
    \Hmat^{[n,m]}\psi^{[m]},
    \qquad n\in\mathcal{R},
\]
where \(\mathcal{R}\) is the assigned row set. This scheduling changes the
memory layout and the parallel work distribution, but it does not change the
Hamiltonian or omit any block products.

A reference pseudocode implementation is:
\begin{lstlisting}
for R_out in output_block_row_groups:

    phi_R = zeros_for_output_rows(R_out)

    for n in R_out:
        for m in range(N_blk):
            H_nm = generate_load_or_fetch_cached_block(n, m)
            phi_R[n] += apply_block(H_nm, psi[m])

    save_or_return_output_rows(R_out, phi_R)
\end{lstlisting}
For an exact dense calculation, all input-block indices \(m\) are included for
each output row \(n\). If the block graph is known to be sparse, the loop over
\(m\) can be restricted to the nonzero block neighbors. Such a restriction is
exact only when the omitted blocks are truly zero; otherwise it defines a
truncated or stochastic approximation.

\section{Hardware-Adaptive Execution Planning}
\label{sec:adaptive}

The block representation defines how the mathematical operator can be applied.
An efficient implementation must also decide how the work should be scheduled
on the available hardware. We therefore introduce an adaptive execution plan.
The plan is a set of choices made before a run, or before each problem size in
a sweep, based on device memory, number of GPUs, expected arithmetic work, and
user accuracy requirements.

An execution plan may include:
\begin{itemize}
    \item the block count \(N_{\mathrm{blk}}\) and block sizes \(b_n\);
    \item the matrix storage mode: procedural, full-cache, partial-cache, or
    externally loaded;
    \item the cache budget per GPU;
    \item the number of cached Hermitian block pairs;
    \item the GPU grouping strategy;
    \item whether work is distributed over independent parameter values,
    output block rows, or both;
    \item a safety estimate for dense arithmetic work;
    \item spectral and sampling parameters used by propagation or correlation
    routines.
\end{itemize}

\subsection{Generic dense matrix--multi-vector planning problem}

The execution planner is not tied to a Hamiltonian. Its generic computational
object is a dense matrix--multi-vector product
\begin{equation}
    Y = A X ,
    \qquad
    A\in\C^{D_{\mathrm{out}}\times D_{\mathrm{in}}},
    \qquad
    X\in\C^{D_{\mathrm{in}}\times R},
    \qquad
    Y\in\C^{D_{\mathrm{out}}\times R}.
    \label{eq:generic_multivector}
\end{equation}
Here (R) denotes the number of vectors to which the same matrix is applied simultaneously. The ordinary matrix-vector case corresponds to (R=1). A block Hamiltonian application in a tensor-product system--bath model can be viewed as a dense bath operator acting on \(R=d_S\) simultaneous state components, one for each system degree of freedom.

For \(n_{\mathrm{op}}\) dense operators applied in one kernel evaluation, the
leading arithmetic scale is
\begin{equation}
    F_{\mathrm{mv}}
    \approx
    8\,n_{\mathrm{op}}\,
    D_{\mathrm{out}}D_{\mathrm{in}}R ,
    \label{eq:generic_matvec_flops}
\end{equation}
where the factor \(8\) is the usual real-operation count scale for one
complex multiply-add.  The square Hamiltonian-vector case is recovered by
setting \(D_{\mathrm{out}}=D_{\mathrm{in}}=D\) and \(R=1\), giving
\(F_{\mathrm{mv}}\sim 8D^2\).  For the kicked TPQ/HMF application considered
below, the dense bath-side operators have
\[
    D_{\mathrm{out}}=D_{\mathrm{in}}=d_B,
    \qquad
    R=d_S,
    \qquad
    n_{\mathrm{op}}=1+n_{\mathrm{int}},
\]
where the first operator is the bath Hamiltonian and the remaining operators
come from the interaction terms.

A generic planning instance is therefore specified by
\[
    \mathcal{P}_{\mathrm{mv}}
    =
    \left(
    D_{\mathrm{out}},
    D_{\mathrm{in}},
    R,
    n_{\mathrm{op}},
    N_{\mathrm{task}},
    B_c,
    M_{\mathrm{free}},
    N_{\mathrm{GPU}}
    \right),
\]
\[
\begin{array}{ll}
D_{\mathrm{out}} & \text{number of output rows of the dense linear map},\\
D_{\mathrm{in}}  & \text{number of input rows, or columns of the dense map},\\
R                & \text{number of simultaneous vectors multiplied}\\
                &\text{by  the same operator},\\
n_{\mathrm{op}}  & \text{number of dense operators applied per kernel call},\\
N_{\mathrm{task}}& \text{number of independent outer-loop tasks},\\
B_c              & \text{bytes per complex scalar},\\
M_{\mathrm{free}}& \text{available accelerator memory},\\
N_{\mathrm{GPU}} & \text{number of available accelerators}.
\end{array}
\]
Thus \(D_{\mathrm{out}}\) and \(D_{\mathrm{in}}\) specify the shape of the
matrix, \(R\) specifies how many vectors are multiplied at once, and
\(n_{\mathrm{op}}\) specifies how many dense matrix actions are fused into
one planned operation.  The task count \(N_{\mathrm{task}}\) counts
independent outer-loop jobs, such as coupling values, disorder realizations,
temperatures, probe vectors, or independent data batches.  The scalar
\(B_c\) is usually \(8\) bytes for single-precision complex arithmetic and
\(16\) bytes for double-precision complex arithmetic.  For a multi-GPU node,
\(M_{\mathrm{free}}\) should be interpreted as the vector
\[
    M_{\mathrm{free}}
    =
    \left(
    M_{\mathrm{free}}^{(1)},\ldots,
    M_{\mathrm{free}}^{(N_{\mathrm{GPU}})}
    \right),
\]
and the conservative memory tests below use its smallest component.  The
planner chooses a block partition, storage mode, cache budget, row-group size,
and distribution of tasks over GPU groups.

\subsection{Memory-constrained cost-model planner}

The first optimizer is an analytic planner.  It enumerates feasible candidate
plans and rejects plans that cannot satisfy a memory safety constraint. If the algorithm keeps (r) temporary arrays of size \(D_{\mathrm{in}}\times R\) in accelerator memory, the workspace memory is estimated as
\begin{equation}
    M_{\mathrm{work}}
    =
    r\,D_{\mathrm{in}}R\,B_c .
    \label{eq:planner_workspace}
\end{equation}
For full caching of \(n_{\mathrm{op}}\) dense matrices over a row group of
size \(p_{\mathrm{row}}\), the per-GPU matrix memory is approximately
\begin{equation}
    M_{\mathrm{full}}
    =
    \frac{
    n_{\mathrm{op}}D_{\mathrm{out}}D_{\mathrm{in}}B_c
    }{
    p_{\mathrm{row}}
    } .
    \label{eq:planner_full_cache}
\end{equation}
For a square Hermitian block representation with equal block size \(b\), a
partial cache containing \(P_{\mathrm{cache}}\) canonical block pairs per
operator costs
\begin{equation}
    M_{\mathrm{partial}}
    =
    n_{\mathrm{op}}P_{\mathrm{cache}}b^2B_c .
    \label{eq:planner_partial_cache}
\end{equation}
The total estimated per-GPU memory is
\begin{equation}
    M_{\mathrm{tot}}
    =
    M_{\mathrm{op}} + M_{\mathrm{work}},
    \qquad
    M_{\mathrm{op}}\in
    \{0,M_{\mathrm{partial}},M_{\mathrm{full}}\},
    \label{eq:planner_total_memory}
\end{equation}
and a plan is accepted only if
\begin{equation}
    M_{\mathrm{tot}}
    \le
    \alpha
    \min_j M_{\mathrm{free}}^{(j)},
    \label{eq:planner_memory_constraint}
\end{equation}
where \(0<\alpha<1\) is a safety fraction.

Among the memory-feasible candidates, the analytic planner assigns a rough
runtime score.  Let \(C(p)\) be the number of GPU groups that can work on
independent tasks under plan \(p\).  The number of task waves is
\begin{equation}
    W(p)
    =
    \left\lceil
    \frac{N_{\mathrm{task}}}{C(p)}
    \right\rceil .
    \label{eq:planner_waves}
\end{equation}
The cost-model score is then
\begin{equation}
    S_{\mathrm{model}}(p)
    =
    W(p)
    \frac{F_{\mathrm{mv}}}{R_{\mathrm{eff}}(p)}
    \Pi_{\mathrm{cache}}(p)
    \Pi_{\mathrm{mem}}(p),
    \label{eq:planner_cost_score}
\end{equation}
where \(R_{\mathrm{eff}}(p)\) is an effective throughput estimate,
\(\Pi_{\mathrm{cache}}\) penalizes procedural regeneration or incomplete
caching, and \(\Pi_{\mathrm{mem}}\) penalizes operation close to the memory
limit.  This score is used only to rank and prune candidates; measured
autotuning, when enabled, overrides it.

\paragraph{Algorithm 1: analytic memory-constrained planning.}
Given \(\mathcal{P}_{\mathrm{mv}}\), enumerate candidate block counts,
operator-storage modes, row-group sizes, and task-grouping modes.  For each
candidate \(p\), compute \(M_{\mathrm{work}}\), \(M_{\mathrm{op}}\), and
\(M_{\mathrm{tot}}\).  Reject \(p\) if Eq.~\eqref{eq:planner_memory_constraint}
fails.  For each surviving plan, compute \(S_{\mathrm{model}}(p)\) from
Eq.~\eqref{eq:planner_cost_score}.  The analytic plan is
\[
    p_{\mathrm{model}}
    =
    \arg\min_p S_{\mathrm{model}}(p).
\]

\subsection{Measured Microbenchmark Autotuning}

The second optimizer is a measured autotuner.  The analytic planner described
above is useful for rejecting plans that cannot fit in memory.  However, it
cannot predict the actual runtime accurately in all cases.  The real runtime
depends on hardware details such as GPU model, memory bandwidth, FP32 or FP64
throughput, cache behavior, communication, kernel launch
overhead, and the cost of generating matrix blocks procedurally.

Therefore, after the analytic planner has produced a small set of feasible
candidate plans, the measured autotuner tests these plans directly on the
available hardware.  Each candidate plan is run using the same matrix-free
kernel that will be used in the full production calculation.  The purpose of
this stage is simple: instead of guessing which feasible plan is fastest, the
code measures it.

Let \(p\) denote one candidate execution plan.  A plan \(p\) includes choices
such as the block size, cache mode, row-group size, GPU grouping, and whether
the calculation is distributed over parameter values, output block rows, or a
hybrid of both.

Before timing the candidate plan, the autotuner performs
\(R_{\mathrm{warm}}\) warmup applications.  These warmup applications are not
included in the timing.  They are used to remove one-time overheads such as
GPU initialization, memory allocation, cache setup, and kernel compilation.

After the warmup stage, the autotuner performs \(R_{\mathrm{bench}}\) timed
applications of a representative matrix-free operation.  In this context, a
representative operation means one application of the same type of
matrix-vector or matrix--multi-vector product that appears in the full
calculation.  It has the same matrix dimensions, block structure, block source,
cache strategy, and GPU distribution as the candidate plan being tested.

If \(t_q(p)\) is the measured time of the \(q\)-th timed application under
plan \(p\), then the average measured time per matrix-free application is
\begin{equation}
    \widehat{t}_{\mathrm{mv}}(p)
    =
    \frac{1}{R_{\mathrm{bench}}}
    \sum_{q=1}^{R_{\mathrm{bench}}}
    t_q(p).
    \label{eq:measured_matvec_time}
\end{equation}
Here:
\begin{itemize}
    \item \(p\) is the candidate execution plan being tested.
    \item \(R_{\mathrm{warm}}\) is the number of warmup applications.
    \item \(R_{\mathrm{bench}}\) is the number of timed benchmark applications.
    \item \(q\) labels one timed benchmark repetition.
    \item \(t_q(p)\) is the measured time of benchmark repetition \(q\) using
    plan \(p\).
    \item \(\widehat{t}_{\mathrm{mv}}(p)\) is the average measured time for
    one matrix-vector or matrix--multi-vector application using plan \(p\).
\end{itemize}

The fastest single matrix-vector application is not always the fastest full
job.  Some plans use all GPUs together for one task, while other plans split
the GPUs into several groups and process independent tasks in parallel.  For
example, if the calculation must be repeated for several coupling strengths,
temperatures, disorder realizations, or random vectors, then different GPU
groups may process different tasks at the same time.

To account for this, the benchmark score includes the number of task waves.
Let \(W(p)\) be the number of waves required to complete all independent tasks
under plan \(p\).  For example, if there are \(N_{\mathrm{task}}\) independent
tasks and plan \(p\) can process \(C(p)\) tasks at the same time, then
\[
    W(p)
    =
    \left\lceil
    \frac{N_{\mathrm{task}}}{C(p)}
    \right\rceil .
\]
Thus, \(W(p)=1\) means that all tasks can be processed in one wave, while
\(W(p)>1\) means that the tasks must be processed in several consecutive
batches.

The total benchmark score of plan \(p\) is defined as
\begin{equation}
    S_{\mathrm{bench}}(p)
    =
    W(p)\widehat{t}_{\mathrm{mv}}(p).
    \label{eq:bench_score}
\end{equation}
Here:
\begin{itemize}
    \item \(S_{\mathrm{bench}}(p)\) is the measured job-level score of plan
    \(p\).
    \item \(W(p)\) is the number of task waves required by plan \(p\).
    \item \(\widehat{t}_{\mathrm{mv}}(p)\) is the measured average time for
    one matrix-free application under plan \(p\).
\end{itemize}
A smaller value of \(S_{\mathrm{bench}}(p)\) means a better plan.

The measured autotuner does not test every possible plan.  It only tests a
small set of promising candidates selected by the analytic planner.  This set
is denoted by \(\mathcal{C}_{\mathrm{bench}}\).  The final measured plan is
chosen as
\begin{equation}
    p_{\mathrm{bench}}
    =
    \arg\min_{p\in\mathcal{C}_{\mathrm{bench}}}
    S_{\mathrm{bench}}(p).
    \label{eq:bench_argmin}
\end{equation}
Here:
\begin{itemize}
    \item \(\mathcal{C}_{\mathrm{bench}}\) is the set of candidate plans tested
    by the measured autotuner.
    \item \(p_{\mathrm{bench}}\) is the plan with the lowest measured
    benchmark score.
\end{itemize}

In summary, the analytic planner first removes impossible or clearly poor
plans.  The measured autotuner then runs a small number of feasible plans on
the actual hardware and selects the one with the best measured job-level
runtime.  This makes the final plan depend on real GPU performance rather than
only on an approximate cost model.

The 18-qubit autotuning result in Fig.~\ref{fig:adaptive-autotune-plan} gives a concrete example: plans with nearly identical single-kernel timings can differ at the job level because they require different numbers of GPU work waves.

\paragraph{Algorithm 2: measured autotuning.}
First run Algorithm 1 and retain the \(K_{\mathrm{bench}}\) best feasible
candidates.  For each candidate, construct the same block source, cache mode,
and GPU grouping that would be used in the production calculation.  Apply the
matrix-free operation several times, synchronizing the GPUs before and after
the timed region.  Record \(\widehat{t}_{\mathrm{mv}}\), memory metadata, GPU
metadata, and the complete plan description.  Choose the candidate minimizing
Eq.~\eqref{eq:bench_score}.

\subsection{Neural surrogate for community-scale plan selection}

The third optimizer is a neural surrogate trained from accumulated benchmark
records.  Each measured trial produces a pair
\[
    \left(x(p),\,\widehat{t}_{\mathrm{mv}}(p)\right),
\]
where \(x(p)\) is a numerical feature vector describing the problem, the plan,
and the hardware.  Typical features include
\[
\begin{aligned}
x(p)=(&
\log D_{\mathrm{in}},
\log D_{\mathrm{out}},
\log R,
\log N_{\mathrm{blk}},
\log b,
\log N_{\mathrm{task}},
p_{\mathrm{row}},\\
W(p),
&\mathbf{1}_{\mathrm{full}},
\mathbf{1}_{\mathrm{partial}},
\mathbf{1}_{\mathrm{procedural}},
\mathbf{1}_{g\text{-parallel}},
\mathbf{1}_{\mathrm{row}},
\mathbf{1}_{\mathrm{hybrid}},\\
&\log F_{\mathrm{mv}},
M_{\mathrm{op}}/M_{\mathrm{free}},
M_{\mathrm{work}}/M_{\mathrm{free}},
M_{\mathrm{tot}}/M_{\mathrm{free}},\\
&\text{GPU memory},
\text{GPU multiprocessors}) .
\end{aligned}
\]
The target is the logarithm of the measured time,
\[
    y(p)
    =
    \log \widehat{t}_{\mathrm{mv}}(p).
\]
The feature vector is normalized componentwise,
\[
    z
    =
    \frac{x-\mu}{\sigma},
\]
and a one-hidden-layer surrogate is
\begin{equation}
    f_{\theta}(x)
    =
    \tanh\!\left(zW_1+b_1\right)W_2+b_2,
    \label{eq:nn_surrogate}
\end{equation}
where
\[
    \theta=\{W_1,b_1,W_2,b_2\}.
\]
The predicted runtime is
\begin{equation}
    \widehat{t}_{\mathrm{NN}}(p)
    =
    \exp(f_{\theta}(x(p))).
    \label{eq:nn_pred_time}
\end{equation}
The network is trained by minimizing the log-time mean-squared error
\begin{equation}
    \mathcal{L}(\theta)
    =
    \frac{1}{N}
    \sum_{i=1}^{N}
    \left[
    f_{\theta}(x_i)
    -
    \log \widehat{t}_{\mathrm{mv},i}
    \right]^2 .
    \label{eq:nn_loss}
\end{equation}
The logarithmic target is used because runtimes are positive and can vary by
orders of magnitude.  The hyperbolic tangent nonlinearity allows the surrogate
to learn interactions between plan variables, such as the fact that a large
row group is beneficial only for compatible cache layouts, or that memory
pressure affects full-cache and procedural modes differently.

The neural surrogate is not used as an unchecked replacement for the physical
or numerical model.  It learns only the empirical hardware map
\[
    \text{problem}+\text{plan}+\text{hardware}
    \longmapsto
    \text{runtime}.
\]
In a hybrid planner, the neural model ranks many feasible candidates, and the
measured autotuner benchmarks only the top \(K_{\mathrm{NN}}\) candidates.
Thus community and industry benchmark data can reduce the search cost on new
machines while a local measurement remains the final decision rule.

\paragraph{Algorithm 3: learned and hybrid plan selection.}
Collect benchmark records from Algorithm 2 in a portable log containing
hardware metadata, plan features, and measured matrix--multi-vector time.
Train \(f_\theta\) using Eq.~\eqref{eq:nn_loss}.  On a new machine, enumerate
memory-feasible candidates with Algorithm 1, rank them by
\(\widehat{t}_{\mathrm{NN}}(p)W(p)\), and either choose the best neural
candidate directly or benchmark the top \(K_{\mathrm{NN}}\) candidates and
select the best measured one using Eq.~\eqref{eq:bench_argmin}.

\subsection{Adaptive block size}

For equal block size \(b\), one dense block requires
\[
    M_{\mathrm{block}} = B_c b^2
\]
bytes, where \(B_c\) is the number of bytes per complex number. Given a target
single-block memory \(M_{\mathrm{target}}\), a natural block size is
\[
    b_{\mathrm{target}}
    \approx
    \sqrt{\frac{M_{\mathrm{target}}}{B_c}} .
\]
The number of blocks can then be chosen so that
\[
    N_{\mathrm{blk}}
    \gtrsim
    \frac{D}{b_{\mathrm{target}}},
\]
subject to divisibility, row-distribution requirements, and model-specific
constraints. In power-of-two Hilbert spaces, choosing \(N_{\mathrm{blk}}\) as
a power of two is often convenient.

The block number should not be interpreted as a physical parameter. It is a
memory-layout and scheduling parameter. If the procedural source is seeded,
partition invariance requires that the generated matrix be independent of the
chosen block partition. Otherwise different block counts correspond to
different random samples and should not be compared as the same Hamiltonian.

\subsection{Adaptive cache selection}

Let \(M_{\mathrm{free}}^{(j)}\) be the free memory on GPU \(j\). A cache budget
can be chosen as
\[
    M_{\mathrm{cache}}
    =
    f_{\mathrm{cache}}
    \min_j M_{\mathrm{free}}^{(j)},
\]
where \(0<f_{\mathrm{cache}}<1\) is a safety fraction. If full caching of the
Hamiltonian blocks fits within this budget, it is usually preferred for
repeated Hamiltonian applications. If not, a partial cache can be used. For a
Hermitian block matrix with equal block size, the number of canonical block
pairs is
\[
    N_{\mathrm{pair}}
    =
    \frac{N_{\mathrm{blk}}(N_{\mathrm{blk}}+1)}{2} .
\]
An approximate number of cached canonical pairs is
\begin{equation}
    P_{\mathrm{cache}}
    =
    \min\left[
    N_{\mathrm{pair}},
    \left\lfloor
    \frac{M_{\mathrm{cache}}}
    {B_c b^2}
    \right\rfloor
    \right].
    \label{eq:partial_cache_pairs}
\end{equation}
The uncached pairs remain procedural or externally loaded. This creates a
smooth transition between no caching and full caching.

Caching is most valuable when the same Hamiltonian is applied many times. Let
\(R_H\) denote the approximate number of Hamiltonian-vector products in a run.
For polynomial real-time propagation,
\[
    R_H \sim N_t p_{\mathrm{rt}},
\]
where \(N_t\) is the number of time samples and \(p_{\mathrm{rt}}\) is the
polynomial order.
The measured growth of \(p_{\mathrm{rt}}\) with Hilbert-space size in the 18-qubit series is shown in Fig.~\ref{fig:adaptive-chebyshev-order}.
More generally, let \(R_A\) denote the approximate number of times the same
matrix or operator \(A\) is applied during a complete run. In repeated
operator-application algorithms one may write schematically
\[
    R_A
    \sim
    \sum_{\ell} N_\ell p_\ell ,
\]
where \(N_\ell\) is the number of repetitions of stage \(\ell\), and
\(p_\ell\) is the number of matrix-vector applications required per repetition.
The stages may represent time stepping, polynomial filtering, Krylov
projection, iterative linear solvers, stochastic trace estimation, optimization
iterations, sampling over random vectors, or repeated applications inside a
machine-learning or scientific-computing pipeline.

When \(R_A\) is large, repeatedly regenerating or loading the same dense blocks
can dominate runtime. In this regime, full GPU-resident caching, or
row-distributed full caching, is strongly preferred whenever memory permits.
Partial caching is useful only when the cached fraction is large enough to
reduce repeated source access substantially.

\subsection{Adaptive parallelization}

Many quantum simulations include independent outer-loop parameters, such as
coupling strengths, disorder realizations, temperatures and more. These can be distributed across GPUs without
communication. We call this parameter-parallel execution.

For a single large Hamiltonian application, the work can instead be split by
output block rows. If GPUs in a group are assigned disjoint row sets
\(\mathcal{R}_j\), GPU \(j\) computes
\[
    \phi^{[n]}
    =
    \sum_m \Hmat^{[n,m]}\psi^{[m]},
    \qquad n\in\mathcal{R}_j .
\]
This row-parallel strategy reduces per-GPU cached matrix memory because each
GPU stores only the block rows it owns. A hybrid strategy combines
parameter-parallelism between GPU groups with row-parallelism inside each
group.

\subsection{Dense-work safety estimates}

For dense block graphs, matrix-free execution reduces matrix-storage pressure
but does not remove dense arithmetic. A useful execution plan should therefore
estimate the cost of one Hamiltonian application. For a dense
\(D\times D\) Hamiltonian, the leading work is proportional to
\[
    W_{H\psi}\sim O(D^2).
\]
In a block implementation with equal block size \(b\), the same estimate can
be written as
\[
    W_{H\psi}
    \sim
    N_{\mathrm{blk}}^2\,O(b^2)
    =
    O(D^2).
\]
If this estimate exceeds a user-defined threshold, the run can be rejected,
warned, or forced explicitly. This prevents a memory-feasible but
time-impractical dense calculation from being submitted accidentally.

For repeated-application algorithms, the more relevant estimate is
\[
    W_{\mathrm{run}}
    \sim
    R_H W_{H\psi},
\]
where \(R_H\) is the number of Hamiltonian applications implied by the
propagator, filter, or correlation routine.

\section{Memory and Runtime Scaling}
\label{sec:memory}

We now estimate the matrix memory footprint and runtime cost of the procedural
block matrix-vector product. For clarity, assume equal block sizes,
\[
    b_n=b,
    \qquad
    D=N_{\mathrm{blk}}b .
\]
Thus \(D\) is the full Hilbert-space dimension, \(N_{\mathrm{blk}}\) is the
number of vector blocks, and each block contains \(b\) complex entries.

A dense explicit Hamiltonian requires storage for a \(D\times D\) complex
matrix,
\begin{equation}
    M_{\mathrm{explicit}}
    =
    B_cD^2
    =
    B_cN_{\mathrm{blk}}^2b^2,
\end{equation}
where \(B_c\) is the number of bytes per complex number. This is the main
memory barrier addressed by the framework.

The block representation does not change the mathematical size of a fully
dense Hamiltonian. If all blocks \(\Hmat^{[n,m]}\) are stored at once on one
device, then the matrix memory is still
\begin{equation}
    M_{\mathrm{blocks}}
    =
    B_cN_{\mathrm{blk}}^2b^2
    =
    B_cD^2 .
\end{equation}
The advantage of the procedural method is that the full set of blocks does
not need to be resident on a single GPU simultaneously. Blocks are generated,
loaded, cached, distributed, or reconstructed only when their action is
required.

Let \(q\) be the number of dense Hamiltonian blocks resident on a GPU at one
time, with
\[
    1\leq q\leq N_{\mathrm{blk}}^2 .
\]
Then the active matrix memory on that GPU is
\begin{equation}
    M_{\mathrm{op,GPU}}
    =
    q B_c b^2 .
    \label{eq:gpumem_dense_q}
\end{equation}
The case \(q=N_{\mathrm{blk}}^2\) recovers full dense matrix storage on one
GPU. The procedural case of interest is \(q\ll N_{\mathrm{blk}}^2\), where
only a limited number of Hamiltonian blocks is materialized, applied, and then
released. In row-distributed full caching, the full matrix is stored across a
GPU group rather than on one device.

For a dense exact matrix-vector product, every output block receives
contributions from every input block:
\begin{equation}
    \phi^{[n]}
    =
    \sum_{m=0}^{N_{\mathrm{blk}}-1}
    \Hmat^{[n,m]}\psi^{[m]} .
    \label{eq:dense_all_blocks_scaling}
\end{equation}
The arithmetic cost remains
\[
    O(D^2).
\]
The procedural method primarily reduces GPU-resident matrix memory. Runtime
improvements occur when repeated source access is avoided by caching, when the
block action is fused efficiently, when row distribution enables larger
matrix-multiplication kernels, or when the Hamiltonian has exact block
sparsity.

\subsection{Runtime cost of procedural block access}

Each block contribution has two costs: the cost of obtaining the block and the
cost of applying it to the input vector block. Let \(t_{\mathrm{src}}(b)\) be
the time required to obtain one \(b\times b\) block, and let
\(t_{\mathrm{mv}}(b)\) be the time required to multiply that block by a
vector block. For the dense exact product,
\[
    N_{\mathrm{blk}}^2
\]
block pairs are included, so a simple runtime model is
\begin{equation}
    T_{H\psi}
    \approx
    N_{\mathrm{blk}}^2
    \left[
        t_{\mathrm{src}}(b)
        +
        t_{\mathrm{mv}}(b)
    \right]
    \label{eq:runtime_no_overlap}
\end{equation}
when source access and computation are not overlapped. If block access and
computation are overlapped by asynchronous execution or double buffering, the
block-pair time is closer to
\begin{equation}
    T_{H\psi}
    \approx
    N_{\mathrm{blk}}^2
    \max
    \left[
        t_{\mathrm{src}}(b),
        t_{\mathrm{mv}}(b)
    \right].
    \label{eq:runtime_overlap}
\end{equation}
Thus, the calculation can be compute-limited, source-limited, or
transfer-limited depending on how the blocks are obtained.

A dense \(b\times b\) block contains \(b^2\) complex entries, so its size is
\begin{equation}
    S_{\mathrm{block}}
    =
    B_c b^2 .
\end{equation}

\paragraph{Externally stored blocks.}
If a block is loaded from file, host memory, a database, or another external
storage layer, then the source time per block is approximately
\begin{equation}
    t_{\mathrm{src}}^{\mathrm{ext}}(b)
    \approx
    t_{\mathrm{lat}}
    +
    \frac{B_cb^2}{\mathcal{B}_{\mathrm{eff}}} .
    \label{eq:db_source_time}
\end{equation}
Here \(t_{\mathrm{lat}}\) is the per-block latency or lookup overhead, and
\(\mathcal{B}_{\mathrm{eff}}\) is the effective bandwidth from the source to
GPU-usable memory.

\paragraph{Seeded generation.}
If a block is generated from a seed, the transferred metadata may be much
smaller than the dense block itself. Let \(S_{\mathrm{seed}}\) be the metadata
size per block and let \(R_{\mathrm{gen}}\) be the effective rate for
generating Hamiltonian entries. Then
\begin{equation}
    t_{\mathrm{src}}^{\mathrm{seed}}(b)
    \approx
    t_{\mathrm{lat}}
    +
    \frac{S_{\mathrm{seed}}}{\mathcal{B}_{\mathrm{meta}}}
    +
    \frac{b^2}{R_{\mathrm{gen}}} .
    \label{eq:seed_source_time}
\end{equation}
Seeded generation reduces storage and data movement because only compact
metadata must be retained. The cost is that the matrix entries must be
regenerated unless the block is cached.

Seeded generation is most attractive when the block is used only a small
number of times, when generation is fused with the contraction, or when memory
is insufficient for caching. It is less attractive when the same block is
needed in many Hamiltonian applications. In repeated propagation or TPQ
filtering, repeated seeded generation can become a dominant cost unless it is
overlapped with computation or replaced by caching.

\paragraph{Analytic generation.}
If a block is defined by an analytic formula
\[
    \Hmat^{[n,m]}_{\alpha\beta}
    =
    f(n,m,\alpha,\beta;\theta),
\]
then the source time depends on the cost of evaluating \(f\). If
\(R_{\mathrm{ana}}\) is the effective rate for evaluating analytic matrix
entries, then
\begin{equation}
    t_{\mathrm{src}}^{\mathrm{ana}}(b)
    \approx
    \frac{b^2}{R_{\mathrm{ana}}} .
    \label{eq:analytic_source_time}
\end{equation}

\paragraph{Cached blocks.}
For cached blocks, the source cost is replaced by cache-access cost. If the
cache is GPU-resident, \(t_{\mathrm{src}}(b)\) can be much smaller than
regeneration or external loading. The price is persistent memory use. Partial
caching makes \(t_{\mathrm{src}}\) block dependent:
\[
    t_{\mathrm{src}}(n,m)
    =
    \begin{cases}
    t_{\mathrm{cache}}, & (n,m)\in \mathcal{P}_{\mathrm{cache}},\\
    t_{\mathrm{regen/load}}, & (n,m)\notin \mathcal{P}_{\mathrm{cache}}.
    \end{cases}
\]

For dense repeated-application workloads, the preferred storage hierarchy is
\begin{align*}
&\text{single-GPU full cache}
\rightarrow \text{row-distributed full cache}\\
&\rightarrow \text{partial cache}
\rightarrow \text{procedural or external streaming}.
\end{align*}
The first two regimes preserve reuse of matrix data. Partial caching reduces
memory but may still require repeated generation or loading of uncached
blocks. Procedural generation minimizes memory but can be much slower when the
same Hamiltonian is applied many times.

\subsection{Selected block graphs}

A secondary case occurs when only a subset of block products is included. This
may happen for an exactly block-sparse Hamiltonian, a controlled truncation,
or a stochastic sampling scheme. If each output block is coupled to only \(z\)
input blocks on average, with \(z\ll N_{\mathrm{blk}}\), then
\begin{equation}
    \phi^{[n]}
    =
    \sum_{m\in\mathcal{N}(n)}
    \Hmat^{[n,m]}\psi^{[m]},
    \qquad
    |\mathcal{N}(n)|\approx z .
\end{equation}
The number of block products is then \(zN_{\mathrm{blk}}\) instead of
\(N_{\mathrm{blk}}^2\). This reduction is exact only if the omitted blocks are
truly zero. If blocks are omitted by truncation or sampling, the result is an
approximate or stochastic matrix-vector product.

\begin{table*}[t]
\centering
\small
\begin{ruledtabular}
\begin{tabular}{lll}
\textbf{Method} & \textbf{GPU-resident matrix memory} & \textbf{Main runtime contribution} \\
\begin{minipage}[t]{0.29\textwidth}Explicit dense global \(\Hmat\)\end{minipage} &
\begin{minipage}[t]{0.25\textwidth}\(O(D^2)\)\end{minipage} &
\begin{minipage}[t]{0.35\textwidth}Dense matrix-vector product after the matrix is resident\end{minipage} \\
\begin{minipage}[t]{0.29\textwidth}Procedural dense matrix, all block pairs included\end{minipage} &
\begin{minipage}[t]{0.25\textwidth}\(O(qb^2)\), with \(1\leq q\leq N_{\mathrm{blk}}^2\)\end{minipage} &
\begin{minipage}[t]{0.35\textwidth}\(N_{\mathrm{blk}}^2[t_{\mathrm{src}}(b)+t_{\mathrm{mv}}(b)]\), or the overlapped maximum\end{minipage} \\
\begin{minipage}[t]{0.29\textwidth}Partially cached block matrix\end{minipage} &
\begin{minipage}[t]{0.25\textwidth}\(O(P_{\mathrm{cache}}b^2)\) plus active workspace\end{minipage} &
\begin{minipage}[t]{0.35\textwidth}Cached blocks use \(t_{\mathrm{cache}}\); uncached blocks use regeneration or loading\end{minipage} \\
\begin{minipage}[t]{0.29\textwidth}Row-distributed full cache\end{minipage} &
\begin{minipage}[t]{0.25\textwidth}Approximately \(O(D^2/p_{\mathrm{row}})\) per GPU\end{minipage} &
\begin{minipage}[t]{0.35\textwidth}Dense row-slice products plus row communication\end{minipage} \\
\begin{minipage}[t]{0.29\textwidth}Selected block graph with \(z\) input neighbors per output block\end{minipage} &
\begin{minipage}[t]{0.25\textwidth}\(O(qb^2)\), with \(q\leq zN_{\mathrm{blk}}\)\end{minipage} &
\begin{minipage}[t]{0.35\textwidth}\(zN_{\mathrm{blk}}[t_{\mathrm{src}}(b)+t_{\mathrm{mv}}(b)]\)\end{minipage}
\end{tabular}
\end{ruledtabular}
\caption{Scaling regimes for \(D=N_{\mathrm{blk}}b\). The dense procedural
case includes all \(N_{\mathrm{blk}}^2\) block products and primarily reduces
GPU-resident matrix memory. Caching changes the source cost but uses
persistent memory. Sparse, truncated, or sampled block graphs reduce the
number of block products only when their assumptions are valid.}
\label{tab:scaling}
\end{table*}

\section{GPU-Resident Matrix Caching and Row Distribution}
\label{sec:gpu}

The most efficient regime for repeated dense Hamiltonian application is to
keep reusable matrix data on the GPU. If the complete dense Hamiltonian cache
fits on one device, the implementation can apply the matrix with large dense
matrix-vector or matrix-matrix kernels. If the cache does not fit on one GPU
but does fit across a group of GPUs, the matrix can be distributed by output
rows.

In row-distributed full caching, a GPU group of size \(p_{\mathrm{row}}\)
stores disjoint row slices of the dense Hamiltonian. GPU \(j\) stores
\[
    H_{\mathrm{local}}^{(j)}
    \in
    \mathbb{C}^{D_j\times D},
\]
where \(D_j\approx D/p_{\mathrm{row}}\). The per-GPU matrix memory is
approximately
\[
    M_{\mathrm{local}}
    \approx
    \frac{B_cD^2}{p_{\mathrm{row}}} .
\]
Each GPU computes its local output segment,
\[
    \phi^{(j)}
    =
    H_{\mathrm{local}}^{(j)}\psi .
\]
The local outputs can then be gathered or kept distributed for subsequent
kernels.

This strategy is useful because the communication volume associated with a
Hamiltonian application is linear in \(D\), while the dense arithmetic is
quadratic in \(D\). For sufficiently large dense matrices, the computation can
therefore remain dominated by local dense linear algebra rather than by the
cost of communicating the vector. The preferred data layout stores each local
row slice contiguously so that the dominant operation is a large GEMM(General Matrix Matrix Multiplication)-like
kernel rather than a long sequence of small block contractions.

Host memory or external storage can be useful as a staging layer, but using it
as the active source for every Hamiltonian application is usually unfavorable
when the same Hamiltonian is applied many times. If matrix data of size
\(M_{\mathrm{op}}\) are streamed from host to GPU for every Hamiltonian-vector
product, the total data movement scales as
\[
    R_H M_{\mathrm{op}},
\]
where \(R_H\) is the number of Hamiltonian applications. For polynomial
propagation, TPQ filtering, and correlation calculations, \(R_H\) can be very
large. In such cases, host-resident matrix data should be transferred to the
GPU once when possible, or used to build a GPU-resident cache. Streaming from
host memory in the inner loop is mainly a fallback for cases where no
GPU-resident representation is feasible.

A practical adaptive implementation therefore attempts the following sequence:
\begin{enumerate}
    \item use a single-GPU full cache when the dense matrix fits on one GPU;
    \item use row-distributed full caching when the dense matrix fits across a
    GPU group;
    \item use partial caching when full caching is impossible but a large
    reusable fraction fits;
    \item use procedural generation or external streaming only when caching is
    not feasible.
\end{enumerate}
This hierarchy is not a physical approximation. It is a storage and scheduling
hierarchy for the same mathematical Hamiltonian.

\section{Example: Autotuned Propagation of an 18-Qubit State on Parallel GPUs}
\label{sec:propagation}

The example consists of a real-time evolution, of the state $\psi$ governed by the Schro\"odinger equation:
\begin{equation}
    i\frac{d}{dt}\psi(t)=\Hmat\psi(t),
\end{equation}
with formal solution
\begin{equation}
    \psi(t+\Delta t)
    =
    e^{-i\Delta t\Hmat}\psi(t).
\end{equation}
We illustrate the current implementation with a concrete run produced by
ADGOM.py. (List \ref{table1}) The example uses the same
Hamiltonian action as above, but the execution plan is not fixed by hand. The
code first constructs a dense-matvec planning problem, rejects memory-infeasible
candidates, benchmarks the remaining candidates on the allocated GPUs, and then
uses the measured fastest plan for the TPQ and correlation calculation.

The simulated Hilbert space contains two system qubits and sixteen bath
qubits,
\[
    d_S = 2^2 = 4,
    \qquad
    d_B = 2^{16}=65536 .
\]
The total Hilbert-space dimension is therefore
\[
    D=d_S d_B = 2^{18}=262144 .
\]
A state vector has \(D\) complex amplitudes. This vector can still be stored on
modern GPUs in single or double precision. The dense Hamiltonian, however,
would require storage for a \(D\times D\) complex matrix. For complex64
arithmetic this is approximately
\[
    8D^2
    =
    8(2^{18})^2
    =
    2^{39}\ \mathrm{bytes}
    \approx 512\ \mathrm{GiB},
\]
and for complex128 arithmetic it is approximately \(1\ \mathrm{TiB}\). Thus
the state vector is not the memory bottleneck; explicit storage of the full
system-bath Hamiltonian is.

The Hamiltonian used in the code has the form
\[
    \Hmat
    =
    H_S\otimes I_B
    +
    I_S\otimes H_B
    +
    g\sum_{a=1}^{N_{\mathrm{int}}} A_a\otimes B_a ,
\]
where \(H_B\) and the \(B_a\) are dense bath-space operators represented in
bath blocks. The action of the Hamiltonian on a state
\[
    \psi\in \C^{d_S}\otimes \C^{d_B}
\]
is evaluated without constructing the full \(D\times D\) Hamiltonian. In block
form, the dominant bath-side operation is
\[
    \phi^{[n]}
    =
    \sum_{m=0}^{N_{\mathrm{blk}}-1}
    H_B^{[n,m]}\psi^{[m]},
\]
together with the analogous block actions for the interaction operators
\(B_a\).

The run used one interaction bath operator, so each Hamiltonian application
uses two dense bath-side operators, \(H_B\) and \(B_1\). In complex64
arithmetic, one full bath-side operator has size
\[
    8d_B^2
    =
    8(65536)^2
    =
    32\ \mathrm{GiB}.
\]
The two bath-side operators therefore require \(64\ \mathrm{GiB}\) if stored
on a single GPU. This exceeds the memory of one L40S GPU, but fits when the
operator rows are distributed over pairs of GPUs. The autotuner selected
\[
    N_{\mathrm{blk}}=16,
    \qquad
    b=\frac{d_B}{N_{\mathrm{blk}}}=4096,
\]
so one complex64 block occupies \(4096^2\times 8=128\ \mathrm{MiB}\). Each GPU
in a two-GPU row group stores a contiguous local slice of shape
\[
    32768\times 65536
\]
for each bath-side operator, corresponding to \(16\ \mathrm{GiB}\) per
operator and \(32\ \mathrm{GiB}\) per GPU for the two operators.

The propagation is performed by repeated matrix-free Hamiltonian applications.
The code uses a centered Chebyshev propagation \cite{tal1984accurate}. First, matrix-free Lanczos
iterations estimate spectral bounds \(E_{\min}\) and \(E_{\max}\). The
Hamiltonian is then scaled as
\[
    \widetilde{\Hmat}
    =
    \frac{\Hmat-cI}{R},
    \qquad
    c=\frac{E_{\max}+E_{\min}}{2},
    \qquad
    R=\frac{E_{\max}-E_{\min}}{2}.
\]
The Chebyshev recurrence is
\begin{align}
    q_0 &= \psi,\\
    q_1 &= \widetilde{\Hmat}q_0,\\
    q_{\ell+1} &= 2\widetilde{\Hmat}q_\ell - q_{\ell-1}.
\end{align}
In block form this becomes
\begin{equation}
    q_{\ell+1}^{[n]}
    =
    2
    \sum_{m=0}^{N_{\mathrm{blk}}-1}
    \widetilde{\Hmat}^{[n,m]}q_\ell^{[m]}
    -
    q_{\ell-1}^{[n]} .
\end{equation}
The numerical intensive operation in the recurrence is the repeated application of
\(\widetilde{\Hmat}\), which is precisely the matrix-free block operation described above.

The calculation was run on a single SLURM node with eight NVIDIA L40S GPUs.
Each GPU reported \(44.39\ \mathrm{GiB}\) total memory, \(43.97\ \mathrm{GiB}\)
free memory at job start, \(142\) streaming multiprocessors, and CUDA compute
capability \(8.9\). The submission script requested eight L40S GPUs, four CPU
cores, and \(64\ \mathrm{GB}\) of host memory, loaded CUDA 12.8.1 when
available, and used a lab-local CuPy kernel cache.

The initial autotuning submission, using ADGOM.py (\texttt{adaptive\_dense\_gpu\_optimized\_matvec.py}), was:
\Needspace{36\baselineskip}
{\bf List VII:} \label{table1}
\begin{lstlisting}
[basicstyle=\ttfamily\scriptsize,
    aboveskip=4pt,
    belowskip=4pt]
python -u adaptive_dense_gpu_optimized_matvec.py \
    --num-gpus 8 \
    --log2-sys 2 \
    --log2-bath-list 16 \
    --seed 3 \
    --num-g 20 \
    --max-g-factor 1.0 \
    --beta 0.2 \
    --eps-list 0.1 \
    --K-hmf 50 \
    --K-corr 1 \
    --num-int-ops 1 \
    --n-blocks auto \
    --operator-mode auto \
    --parallel-mode auto \
    --row-group-size auto \
    --planner-mode autotune \
    --autotune-repeats 3 \
    --autotune-warmup 1 \
    --autotune-max-candidates 8 \
    --plan-trial-log q18_bath16_plan_trials.jsonl \
    --sampling-source bare \
    --sampling-spacing-floor 0 \
    --sampling-max-N 200 \
    --dtype complex64 \
    --cheb-tol 1e-7 \
    --lanczos-m 40 \
    --skip-exact-rdm \
    --no-diagnostics \
    --checkpoint-dir q18_bath16_auto_checkpoints
\end{lstlisting}

The measured planner benchmarked eight candidate plans. The selected plan was
full-cache hybrid execution with row-group size \(2\):
\[
    \mathrm{gpu\ groups}
    =
    [[0,1],[2,3],[4,5],[6,7]] .
\]
The selected plan used \(N_{\mathrm{blk}}=16\), \(b=4096\), full-cache storage,
and complex64 arithmetic. The planner estimated \(32\ \mathrm{GiB}\) of
operator storage per GPU, \(116\ \mathrm{MiB}\) of state/workspace memory per
primary GPU, and \(32.1\ \mathrm{GiB}\) total memory per GPU. The estimated
work per Hamiltonian application was
\[
    2\,d_S d_B^2\,8
    =
    2.749\times 10^{11}
\]
real floating-point operations. The measured microbenchmark time was
\(0.104089\ \mathrm{s}\) per \(\Hmat\psi\) application for the selected plan. The
effective planner score was \(0.520447\ \mathrm{s}\), corresponding to five
waves of work for \(20\) coupling values over four GPU groups.

\begin{figure*}[t]
\centering
\includegraphics[width=0.95\textwidth]{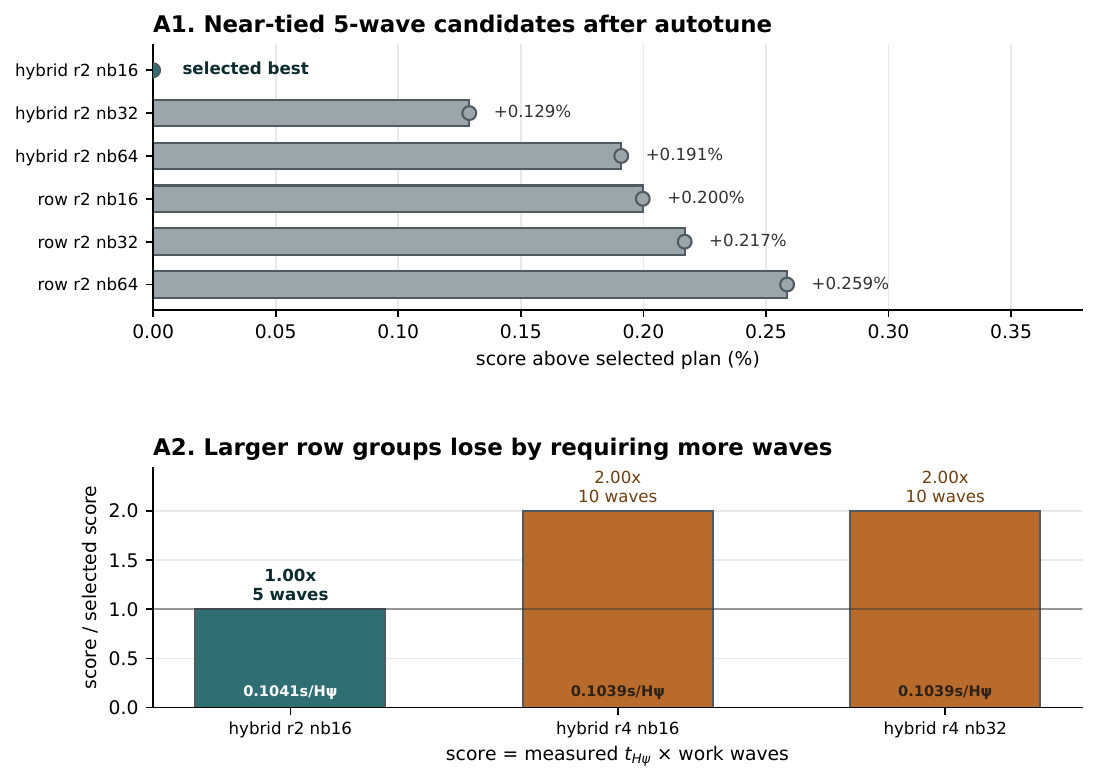}
\caption{Measured autotuning of the 18-qubit execution plan on eight NVIDIA
L40S GPUs. A1 shows the near-tied five-wave candidates after the autotune
stage, plotted as the score increase relative to the selected plan. The
selected full-cache hybrid plan with row-group size \(2\) and
\(N_{\mathrm{blk}}=16\) is only marginally faster than several alternatives,
but it has the lowest measured score. A2 separates the reason that some
apparently competitive candidates are rejected: plans with row-group size
\(4\) have nearly the same measured \(\Hmat\psi\) time, but they require ten
GPU work waves rather than five. The autotuner therefore chooses the plan that
minimizes total scheduled work, not merely the plan with the smallest
 single-application timing.}
\label{fig:adaptive-autotune-plan}
\end{figure*}

The measured autotuning outcome is shown in Fig.~\ref{fig:adaptive-autotune-plan}.
The best few feasible plans are very close in raw
\(\Hmat\psi\) time, so the planner decision is not determined by a single
kernel timing alone. The relevant quantity is the measured kernel time
multiplied by the number of GPU work waves required to cover all coupling
values.

\begin{figure}[t]
\centering
\includegraphics[width=\columnwidth]{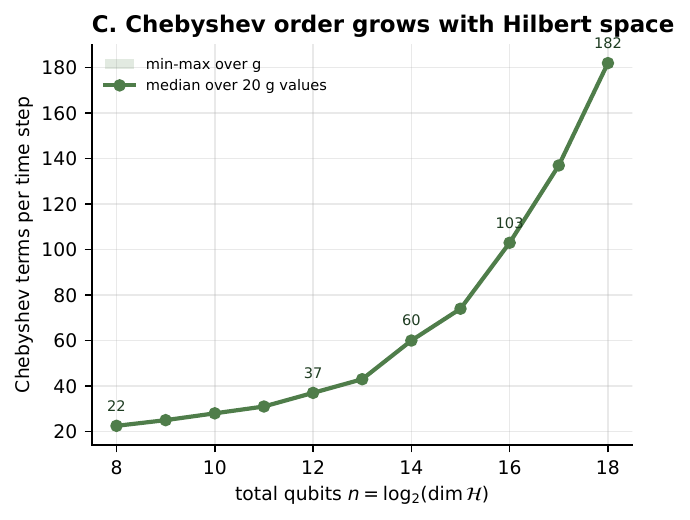}
\caption{Real-time Chebyshev order as a function of Hilbert-space size for the
same completed two-system-qubit calculations. For each total qubit count, the
order was computed from the saved Lanczos half-widths \(R\), the selected time
step \(\tau\), and the same coefficient-tolerance rule used by the propagation
code. Points show the median over the twenty coupling values, and the shaded
band shows the corresponding minimum--maximum range. The increase from about
\(23\) terms at \(2^8\) dimension to about \(182\) terms at \(2^{18}\)
demonstrates that the larger simulations are more expensive not only because
the vectors and block operations are larger, but also because each physical
time step requires more Chebyshev Hamiltonian applications.}
\label{fig:adaptive-chebyshev-order}
\end{figure}

The physical and numerical parameters were
\[
\begin{aligned}
    \beta &= 0.2,\qquad K_{\mathrm{HMF}}=50,\qquad K_{\mathrm{corr}}=1,
    \qquad N=200,\\
    \tau &= 0.1982748544807706 .
\end{aligned}
\]
The coupling grid contained \(20\) values from \(g=0\) to
\(g_{\max}=6.3378539\). The bare-system sampling rule selected
\[
    T=N\tau=39.654971,
    \qquad
    \pi/\tau=15.844635 .
\]
For the completed checkpointed trajectories, matrix-free Lanczos gave spectral
radii near \(713\). The imaginary-time TPQ Chebyshev filter used order \(80\),
and the centered real-time Chebyshev propagator used order \(181\) per time
step.
The dependence of this Chebyshev order on Hilbert-space dimension is shown in
Fig.~\ref{fig:adaptive-chebyshev-order}.

\begin{figure}[t]
\centering
\includegraphics[width=\columnwidth]{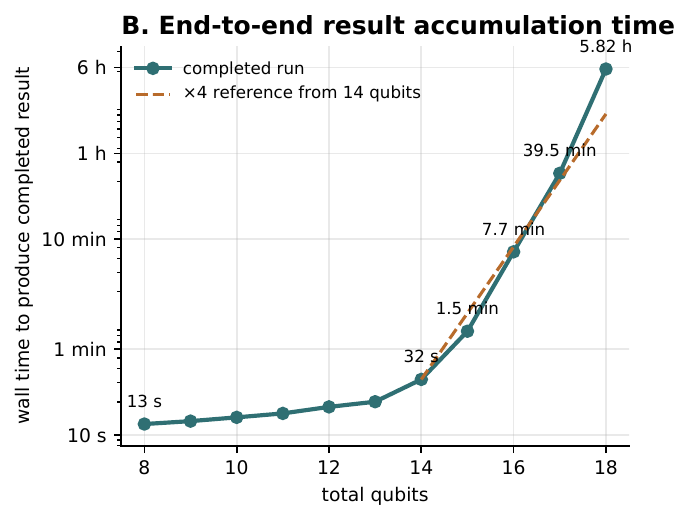}
\caption{End-to-end accumulation time for the completed two-system-qubit
calculations. Each point is a completed run with \(N=200\) time samples,
\(K_{\mathrm{HMF}}=50\), \(K_{\mathrm{corr}}=1\), one interaction bath operator,
and complex64 arithmetic. The dashed line is a reference \(4\times\) growth per
added bath qubit, anchored at the 14-qubit result where the dense bath-side
work begins to dominate. The measured curve shows the practical cost of
producing the saved result files, including propagation, scheduling,
checkpoint handling when used, and final assembly.}
\label{fig:adaptive-wall-time}
\end{figure}

The calculation was completed in two checkpointed SLURM submissions. The
initial autotuning job selected the plan above and wrote checkpoints for the
fifteen zero-based coupling indices \(5,\ldots,19\). The checkpoint
continuation then reused the same directory,
\(\texttt{q18\_bath16\_auto\_checkpoints}\), and specified the selected plan
directly,
\[
    \texttt{full-cache},\quad \texttt{hybrid},\quad
    \texttt{row-group-size}=2,\quad N_{\mathrm{blk}}=16 ,
\]
without repeating the autotuning benchmark. This continuation completed the
missing zero-based indices \(0,\ldots,4\), so the checkpoint directory now
contains
\[
    \texttt{gidx\_0000.npz},\ldots,\texttt{gidx\_0019.npz}.
\]
The continuation job finished in \(6258.74\ \mathrm{s}\) and assembled the
final output file

\[
\begin{aligned}
&\texttt{wave\_adaptive\_dense\_tpq\_raw\_modefull-cache}\\   &\texttt{\_parhybrid\_KHMF50\_KC1\_nint1\_2\_16\_N200}\\   &\texttt{\_tau0p198274854481\_nb16.npz}.
\end{aligned}
\]
Figures~\ref{fig:adaptive-wall-time} and~\ref{fig:adaptive-chebyshev-order}
summarize the computational scaling observed in the same two-system-qubit
series. The end-to-end runtime reports the time needed to produce a completed result
file, while the Chebyshev order reports the number of Hamiltonian applications
required for one real-time step.

The saved arrays had shapes
\[
    \texttt{corr\_per\_g\_eps}=(20,1,200),
    \
    \texttt{C\_per\_tpq}=(20,1,1,200).
\]
Excluding the non coupling case across all nineteen coupling values, the measured propagation times were tightly
clustered between \(3745.79\ \mathrm{s}\) and \(3754.00\ \mathrm{s}\), with
mean \(3749.84\ \mathrm{s}\).
The resulting end-to-end runtime scaling is summarized in
Fig.~\ref{fig:adaptive-wall-time}.
This example demonstrates the practical role of the method. The dense
Hamiltonian action is not made cheap: the calculation remains a large dense
matrix-free propagation. The advantage is that the execution plan is chosen on
the target hardware and the calculation is no longer blocked by the requirement
to assemble or store the full \(2^{18}\times 2^{18}\) Hamiltonian on one GPU.
The memory requirement is converted into a combination of full GPU-resident
row caches, state/workspace buffers, checkpointing, and measured GPU grouping.

\section{Limitations and Design Implications}
\label{sec:limitations}

The framework changes matrix memory scaling and scheduling, but it does not
make a fully dense Hamiltonian cheap. If all block pairs are nonzero, the
exact Hamiltonian-vector product still has dense arithmetic cost. In such a
case, the method can make the calculation possible from the standpoint of
matrix storage, but the remaining runtime may still be impractical.
This separation between memory feasibility and runtime cost is visible in Fig.~\ref{fig:adaptive-wall-time}, where the calculation becomes possible but the completed-result time still grows rapidly with Hilbert-space size.

This limitation leads to several design implications. First, caching is useful
only when the cached data are reused. Repeated propagation and TPQ filtering
are favorable because they apply the same Hamiltonian many times. Second,
partial caching is not automatically efficient. If a large fraction of the
matrix must still be regenerated or loaded at every Hamiltonian application,
partial caching can be much slower than full caching. Third, row-distributed
full caching is preferable when the full dense matrix does not fit on one GPU
but does fit across a GPU group. Fourth, streaming matrix data from CPU memory
in the inner loop should be treated as a fallback, not as the preferred regime,
when the Hamiltonian is applied many times.

Fifth, optimization of the execution plan should be interpreted as empirical
performance modeling, not as a physical approximation. The analytic planner
rejects memory-infeasible candidates and ranks feasible candidates by a cost
model. The measured autotuner replaces that ranking by direct timings on the
target hardware. The neural surrogate learns only the hardware-dependent map
from problem features and plan features to runtime. It should therefore be
used as a ranking prior, or as part of a hybrid planner followed by local
measurement, unless it has been validated on the relevant hardware class and
problem family.

The block partition should also be designed carefully. Too few blocks can make
single-block memory too large and prevent row distribution. Too many blocks can
increase overhead and reduce arithmetic intensity. In dense row-distributed
execution, the best implementation stores local row slices as contiguous
matrices and applies them with large GEMM-like kernels. The mathematical block
partition remains general, but the data layout strongly affects performance.

Finally, computational parameters must be separated from the Hamiltonian being
simulated. The block count, block size, cache mode, GPU grouping, and
row-group size are execution choices. They should not change the Hamiltonian.
If a procedural source depends on the block partition, then these execution
choices become part of the random sample and must be reported and validated. A
partition-invariant source avoids this problem by defining matrix entries from
global indices rather than from block labels.

\section{Discussion}
\label{sec:discussion}

The framework presented here is motivated by a simple principle: in large
quantum simulations, the mathematical Hilbert space may be fixed by the
physics, but the memory representation of the Hamiltonian matrix is an
algorithmic choice. Explicitly storing the full Hamiltonian on one GPU should
not be the only route to exact or controlled quantum-state propagation.

The method separates the mathematical Hamiltonian from its storage. The
operator \(\Hmat\) remains a single well-defined block matrix, but its entries
are supplied procedurally and only when required by the current computation.
This makes it possible to apply Hamiltonians whose full dense representation
would exceed single-GPU memory. In this sense, the framework turns the size of
the matrix from a hard storage constraint into a scheduling, generation,
caching, data-layout, and bandwidth problem.

The block representation in Eq.~\eqref{eq:blockham} is mathematically general:
any finite matrix can be partitioned into blocks. Its computational value comes
from how those blocks are generated, loaded, compressed, fused, cached,
distributed, or scheduled. For a fully dense block matrix, the method primarily
reduces GPU memory requirements rather than arithmetic complexity. If the
block graph is sparse, so that only \(m\in\mathcal{N}(n)\) contribute to each
output block, then the product becomes
\[
    \phi^{[n]}
    =
    \sum_{m\in\mathcal{N}(n)}
    \Hmat^{[n,m]}\psi^{[m]},
\]
which can reduce both memory traffic and arithmetic cost. The direct-sum case
is the limiting situation \(\mathcal{N}(n)=\{n\}\).

Hardware adaptiveness is not a separate physical approximation. It is an
execution strategy for the same mathematical operation. A large-memory GPU may
prefer full caching; a smaller GPU may require row-distributed full caching,
partial caching, or procedural generation; a parameter scan may prefer
independent GPU workers; a single large calculation may prefer row-parallel
execution. The optimal strategy is therefore not universal. It depends on the
relationship between matrix size, cache budget, bandwidth, data layout, and
arithmetic throughput.
The measured plan comparison in Fig.~\ref{fig:adaptive-autotune-plan} is an example of this hardware dependence: the selected plan is determined by the combination of memory feasibility, kernel timing, and available GPU concurrency.
The analytic, measured, and learned planners make this dependency explicit.
The analytic planner gives a reproducible memory-feasible baseline; the
measured autotuner adapts the baseline to the actual accelerator, driver,
interconnect, and kernel implementation; and the neural surrogate provides a
way to reuse benchmark experience across machines. Because the feature vector
is defined for the generic dense matrix--multi-vector product, the same
planner can be applied to the kicked TPQ/HMF model, to other Hamiltonian
families, and to non-Hamiltonian large linear maps. For example the Liouvillian generator ${\cal L}$ in open quantum systems \cite{kosloff2019quantum,PhysRevE.94.022126}.

This data-driven layer also suggests a useful community and industrial
workflow. Each benchmark record contains the mathematical problem size, the
candidate execution plan, hardware metadata, and the measured
matrix--multi-vector time. Such records do not reveal the full scientific
dataset or proprietary matrix entries; they describe how fast a declared
linear-algebra operation ran under a declared plan. Aggregating these records
would allow the neural surrogate to improve as more laboratories, clusters,
and industrial deployments contribute timings.

This distinction is especially important when the matrix is generated
procedurally. Cache mode should not change the Hamiltonian, but the generation
rule can accidentally make the block partition part of the random sample. For
physical comparisons at fixed disorder realization or fixed random matrix
sample, the procedural generator must be partition invariant.

The method is most advantageous when the cost of generating or loading a block
is small compared with the cost of storing the global matrix, when cached
blocks are reused many times, when the block action can be fused with the
contraction, when row-distributed caching enables large local GEMM operations,
or when the same Hamiltonian is applied repeatedly. Propagation, TPQ filtering,
and correlation calculations all have this repeated-application structure, so
implementation effort spent on the matrix-free kernel and adaptive execution
plan is amortized over many Hamiltonian applications.

The general approach can also be adopted to quantum computing using the idea of Block-encoding and unitary matrix dilation
\cite{joven2026scalable}.

\section{Conclusion}

We have presented a procedural matrix-free and hardware-adaptive framework for
applying very large Hamiltonians on GPUs without storing the full dense matrix
on a single device. The Hamiltonian is treated as one block-partitioned
operator,
\[
    \Hmat=\left(\Hmat^{[n,m]}\right)_{n,m=0}^{N_{\mathrm{blk}}-1},
\]
and the matrix-vector product is evaluated by the blockwise rule
\[
    \phi^{[n]}
    =
    \sum_{m=0}^{N_{\mathrm{blk}}-1}
    \Hmat^{[n,m]}\psi^{[m]} .
\]
The global \(D\times D\) Hamiltonian is never assembled on a single device.
Matrix blocks are generated, loaded, cached, distributed, or reconstructed only
when needed.

The main contribution is a memory-scalable architecture for Hamiltonian
application together with an adaptive execution strategy. For a fully dense
Hamiltonian, the arithmetic cost of the exact matrix-vector product is not
removed. However, the GPU-resident matrix-memory requirement is changed
fundamentally: instead of requiring the full Hamiltonian on one GPU, the
computation can require only selected blocks, cached data, row-distributed
matrix slices, and temporary workspace. This allows simulations to be
attempted in regimes where explicit Hamiltonian storage would otherwise be
impossible.

We have also formulated execution planning itself as a general dense
matrix--multi-vector optimization problem. In this formulation, the planner
chooses among mathematically equivalent implementations of \(Y=AX\). The paper
describes three plan-selection algorithms: a memory-constrained analytic
cost-model planner, a measured microbenchmark autotuner, and a neural
surrogate trained from logged trials. The learned model does not alter the
operator or the numerical method; it predicts which feasible execution plan is
likely to run fastest, and in the hybrid mode it reduces the number of local
benchmarks needed before the final measured choice is made.

The same matrix-application kernel supports quantum real-time propagation,
imaginary-time filtering, stochastic TPQ preparation, observable estimation,
and correlation-function evaluation \cite{endo2018linear}. In each of these cases, the essential
computational task is the repeated application of a large linear operator to
one or more vectors. Beyond quantum simulation, the same structure appears in
many algorithms dominated by large linear maps. This includes scientific
machine learning and artificial-intelligence workloads, where matrix-vector
or matrix-matrix operations appear in linear layers, projection operators,
embedding maps, attention mechanisms, kernel methods, iterative optimization,
and large-scale inference \cite{dao2022flashattention}.

The procedural interface used here is therefore not tied to Hamiltonian
dynamics alone. Hamiltonian-vector multiplication is a prominent
quantum-mechanical example of a broader class of memory-limited
operator-vector and matrix-vector multiplication problems. Within this broader
view, seeded random blocks, analytic formulas, fused block actions, cached
blocks, and externally stored deterministic matrix data are simply different
ways of supplying the local pieces of a large linear map without explicitly
materializing the full matrix on a single device.

As exact and controlled quantum simulations become increasingly important for
benchmarking quantum computers and developing quantum technologies, matrix
memory scalability becomes a central requirement. The framework developed here
provides a route toward simulations in which available GPU memory is no longer
identified only with the ability to store the full Hamiltonian matrix. The
remaining limits become explicit and tunable: block generation, caching, data
movement, parallel scheduling, arithmetic throughput, and algorithmic accuracy.

\begin{acknowledgments}
We thank Raam Uzdin for support and useful discussion. We acknowledge the support of the Hebrew University Research Computing Services (HURCS) for hosting this study and The Fritz Haber center for overall support.
\end{acknowledgments}

\bibliography{ref}

@article{saad1986gmres,
  title={GMRES: A generalized minimal residual algorithm for solving nonsymmetric linear systems},
  author={Saad, Youcef and Schultz, Martin H},
  journal={SIAM Journal on scientific and statistical computing},
  volume={7},
  number={3},
  pages={856--869},
  year={1986},
  publisher={SIAM}
}

@article{kosloff1986direct,
  title={A direct relaxation method for calculating eigenfunctions and eigenvalues of the Schr{\"o}dinger equation on a grid},
  author={Kosloff, Ronnie and Tal-Ezer, H},
  journal={Chemical Physics Letters},
  volume={127},
  number={3},
  pages={223--230},
  year={1986},
  publisher={Elsevier}
}

@article{kosloff1994propagation,
  title={Propagation methods for quantum molecular dynamics},
  author={Kosloff, Ronnie},
  journal={Annual review of physical chemistry},
  volume={45},
  number={1},
  pages={145--178},
  year={1994}
}

@article{baer1997chebyshev,
  title={Chebyshev expansion methods for electronic structure calculations on large molecular systems},
  author={Baer, Roi and Head-Gordon, Martin},
  journal={The Journal of chemical physics},
  volume={107},
  number={23},
  pages={10003--10013},
  year={1997},
  publisher={American Institute of Physics}
}

@article{sugiura2013canonical,
  title={Canonical thermal pure quantum state},
  author={Sugiura, Sho and Shimizu, Akira},
  journal={arXiv preprint arXiv:1302.3138},
  year={2013}
}

@article{endo2018linear,
  title={From linear to nonlinear responses of thermal pure quantum states},
  author={Endo, Hiroyuki and Hotta, Chisa and Shimizu, Akira},
  journal={Physical review letters},
  volume={121},
  number={22},
  pages={220601},
  year={2018},
  publisher={APS}
}

@article{wietek2018sublattice,
  title={Sublattice coding algorithm and distributed memory parallelization for large-scale exact diagonalizations of quantum many-body systems},
  author={Wietek, Alexander and L{\"a}uchli, Andreas M},
  journal={Physical Review E},
  volume={98},
  number={3},
  pages={033309},
  year={2018},
  publisher={APS}
}

@article{charlier2021kernel,
  title={Kernel operations on the GPU, with autodiff, without memory overflows},
  author={Charlier, Benjamin and Feydy, Jean and Glaunes, Joan Alexis and Collin, Fran{\c{c}}ois-David and Durif, Ghislain},
  journal={Journal of Machine Learning Research},
  volume={22},
  number={74},
  pages={1--6},
  year={2021}
}

@inproceedings{westerhout2023implementing,
  title={Implementing scalable matrix-vector products for the exact diagonalization methods in quantum many-body physics},
  author={Westerhout, Tom and Chamberlain, Bradford L},
  booktitle={Proceedings of the SC'23 Workshops of the International Conference on High Performance Computing, Network, Storage, and Analysis},
  pages={1140--1150},
  year={2023},
  doi={10.1145/3624062.3624597}
}

@article{tal1984accurate,
  title={An accurate and efficient scheme for propagating the time dependent Schr{\"o}dinger equation},
  author={Tal-Ezer, Hillel and Kosloff, Ronnie},
  journal={The Journal of chemical physics},
  volume={81},
  number={9},
  pages={3967--3971},
  year={1984},
  publisher={American Institute of Physics}
}

@article{wall1995extraction,
  title={Extraction, through filter-diagonalization, of general quantum eigenvalues or classical normal mode frequencies from a small number of residues or a short-time segment of a signal. I. Theory and application to a quantum-dynamics model},
  author={Wall, Michael R and Neuhauser, Daniel},
  journal={The Journal of chemical physics},
  volume={102},
  number={20},
  pages={8011--8022},
  year={1995},
  publisher={American Institute of Physics}
}

@article{eckseler2025escaping,
  title={Escaping the Krylov space during the finite-precision Lanczos algorithm},
  author={Eckseler, Jannis and Pieper, Max and Schnack, J{\"u}rgen},
  journal={Physical Review E},
  volume={112},
  number={2},
  pages={025306},
  year={2025},
  publisher={APS}
}

@article{braun2014numerical,
  title={Numerical evaluation of Green's functions based on the Chebyshev expansion},
  author={Braun, Alexander and Schmitteckert, Peter},
  journal={Physical Review B},
  volume={90},
  number={16},
  pages={165112},
  year={2014},
  publisher={APS}
}

@article{joven2026scalable,
  title={Scalable quantum computational science: A perspective from block-encodings and polynomial transformations},
  author={Joven, Kevin J and Das, Elin Ranjan and Bierman, Joel and Majumdar, Aishwarya and Heris, Masoud Hakimi and Liu, Yuan},
  journal={APL Computational Physics},
  volume={2},
  number={1},
  year={2026},
  publisher={AIP Publishing}
}

@article{saad1989numerical,
  title={Numerical solution of large nonsymmetric eigenvalue problems},
  author={Saad, Youcef},
  journal={Computer Physics Communications},
  volume={53},
  number={1-3},
  pages={71--90},
  year={1989},
  publisher={Elsevier}
}

@article{dax2017new,
  title={A new type of restarted Krylov methods},
  author={Dax, Achiya},
  journal={Adv. Linear Algebra Matrix Theory},
  volume={7},
  pages={18--28},
  year={2017}
}

@article{baglama2005augmented,
  title={Augmented implicitly restarted Lanczos bidiagonalization methods},
  author={Baglama, James and Reichel, Lothar},
  journal={SIAM Journal on Scientific Computing},
  volume={27},
  number={1},
  pages={19--42},
  year={2005},
  publisher={SIAM}
}

@article{de1987product,
  title={Product formula algorithms for solving the time dependent Schr{\"o}dinger equation},
  author={De Raedt, Hans},
  journal={Computer Physics Reports},
  volume={7},
  number={1},
  pages={1--72},
  year={1987},
  publisher={Elsevier}
}

@article{de2026universal,
  title={Universal quantum computer simulation of 50 qubits on Europe’s first exascale supercomputer harnessing its heterogeneous CPU--GPU architecture},
  author={De Raedt, Hans and Kraus, Jiri and Herten, Andreas and Mehta, Vrinda and Bode, Mathis and Hrywniak, Markus and Michielsen, Kristel and Lippert, Thomas},
  journal={Future Generation Computer Systems},
  pages={108592},
  year={2026},
  publisher={Elsevier}
}

@article{PhysRevE.94.022126,
  title = {Dynamics of open quantum spin systems: An assessment of the quantum master equation approach},
  author = {Zhao, P. and De Raedt, H. and Miyashita, S. and Jin, F. and Michielsen, K.},
  journal = {Phys. Rev. E},
  volume = {94},
  issue = {2},
  pages = {022126},
  numpages = {19},
  year = {2016},
  month = {Aug},
  publisher = {American Physical Society},
  doi = {10.1103/PhysRevE.94.022126}
}

@article{kosloff2019quantum,
  title={Quantum thermodynamics and open-systems modeling},
  author={Kosloff, Ronnie},
  journal={The Journal of chemical physics},
  volume={150},
  number={20},
  year={2019},
  publisher={AIP Publishing}
}

@inproceedings{dao2022flashattention,
  title={FlashAttention: Fast and Memory-Efficient Exact Attention with IO-Awareness},
  author={Dao, Tri and Fu, Daniel Y. and Ermon, Stefano and Rudra, Atri and R{\'e}, Christopher},
  booktitle={Advances in Neural Information Processing Systems},
  volume={35},
  pages={16344--16359},
  year={2022}
}
\end{document}